\documentclass[12pt]{article}

\font\bfs=cmb10 at 7pt

\font\bb=msbm10 at 10pt

\font\cal=cmsy10 at 9pt
\font\cals=cmsy10 at 8pt

\font\rms=cmr10 at 8pt

\font\sf=cmss10 at 10pt %san serif

\def\0#1{\mbox{\rm#1}}
\def\1#1{\mbox{\bb#1}}
\def\2#1{\mbox{\bf#1}}
\def\3#1{{\cal #1}}
\def\4#1{\mbox{\cals#1}}
\def\5#1{\mbox{\sf#1}}
\def\6#1{\mbox{\rms #1}}
\def\7#1{\mbox{\bfs #1}}
\def\8#1{{\tilde #1}}
\def\9#1{{\breve #1}}

\def\BEq{\begin{equation}}
\def\EEq{\end{equation}}
\def\BEqA{\begin{eqnarray}}
\def\EEqA{\end{eqnarray}}
\def\BEn{\begin{enumerate}}
\def\EEn{\end{enumerate}}

\def\Rbb{\mbox{\bb R}}

\def\tav{\hbox{
\kern-1pt\rule[0pt]{1.5pt}{.8pt}{\kern-3.3pt}
\rule[0pt]{.4pt}{5pt}{\kern-4.4pt}
\rule[5pt]{4.5pt}{.8pt}{\kern-3.45pt}
\rule[0pt]{.4pt}{5.3pt}{\kern-1pt}
}}

\def\from{\kern-2pt\leftarrow\kern-2pt}

\def\II{|\kern-1pt |}

\def\tav{\hbox{
\kern-1.0pt
\rule[0pt]{1.3pt}{.8pt}{\kern-3.6pt}
\rule[0pt]{.4pt}{6pt}{\kern-3.0pt}
\rule[4.5pt]{3.0pt}{.8pt}{\kern-3.3pt}
\rule[0pt]{.4pt}{5pt}{\kern-1pt}
}}

\def\from{\kern-2pt\leftarrow\kern-2pt}

\def\lmult{{\lfloor\kern-5pt\lfloor}}
\def\rmult{{\rfloor\kern-5pt\rfloor}}

\def\II{|\kern-1pt |}

\usepackage{makeidx,graphics,latexsym}

\begin{document}

\title{{\bf Modified Einstein equations}}

\author{
Andrei A.
Galiautdinov\footnote{
e-mail: agaliautdinov@lib.brenau.edu}
\\
\normalsize{\it Brenau University, Gainesville, Georgia
30501}
}

\date{\today}

\maketitle

\abstract{Standard general relativity fails to take into
account the changes in coordinates induced by the variation
of metric in the Hilbert action principle. We
propose to include such changes by introducing a fundamental
compensating tensor field and modifying the usual
variational procedure.}

\section{Introduction}

Mathematical representation of
spacetime in general relativity relies heavily on the
idea that spacetime events form a differentiable manifold.
The manifold topology of spacetime (same as that of
Euclidean four-dimensional space $\Rbb^4$) is postulated
before introduction of any (sufficiently differentiable in
that topology) metric.

Finkelstein had argued that the
assumption that topology is prior to metric is operationally
suspect because experimentally it is always
determined from the exchange of signals, governed by the
metric. Since the assignment of coordinates
to spacetime events is also based on propagation of signals,
the usual point of view that coordinates are independent of
the metric is equally unsound: there are no
coordinates to distinguish one point from another besides
the physical fields themselves. The dependence of spacetime
coordinates on the metric would modify the Hilbert action
principle and, consequently, Einstein's field equations.
This is in contrast to the usual theory which fails to take
into account the changes in the coordinates induced by the
variation in the metric \cite{FINK97}.

By the metric of spacetime I mean the collection of all
possible proper intervals between all possible
spacetime events. The ability of such collections of numbers
to  represent the shapes of differentiable
manifolds led to the invention of
Riemannian geometry.

In general
relativity we mostly deal with two types of metric
variations. Variations due to simple changes in coordinates
({\it coordinate variations}) do not affect the ''shape'' of
spacetime and represent going from one reference
frame to another. In this case the metric transforms in the
usual way as a second rank tensor according to
\BEq
\delta^{\rm c}: \quad x(P) \rightarrow x'(P), \quad g_{\mu
\nu}(x)
\rightarrow g'_{\mu \nu}(x')=\frac{\partial
x^\alpha}{\partial x'^\mu}\frac{\partial x^\beta}{\partial
x'^\nu}g_{\alpha\beta}(x).
\EEq
Variations of the second type (let's call them {\it
functional variations}),
\BEq
\delta^{\rm f}: \quad x(P)={\rm const}, \quad 
g_{\mu \nu}(x) \rightarrow
g'_{\mu \nu}(x)=g_{\mu \nu}(x)+\delta^{\rm f} g_{\mu
\nu}(x),
\EEq
which are used in the action
principle, do not change coordinates but modify the
functional form of the metric. These can be viewed as
changes in the ''shape'' of spacetime. General relativity
actually fails to provide the physical meaning of such
variations, whose backreaction on spacetime
coordinates could become important and would have to be
taken into account.

A question arises whether all these
variations change the identity of spacetime events/points, or
whether the spacetime with all its points ''intact'' can
still be viewed as a stage, albeit the dynamical one, for the
matter (the ''relative space vs. absolute space'' debate
\cite{DICKE64}). In quantum theory of spacetime these
questions would be of paramount importance, because events
and elementary particle processes are fundamentally the
same. For now, however, we can view spacetime points --
marked in one way or another -- as just ''being there,''
even when the metric is varied (similar to how all the points
on a mattress would still be there even when it is
deformed), and speak of the variational relationship 
\BEq
\delta^{\rm f} g_{\mu \nu}
\rightarrow
\delta^{\rm f} x^\mu.
\EEq
This relationship is clearly nonlocal.\footnote{I thank
David Finkelstein for drawing my attention to this
particular point.}

\section{Variational principle modified}

Let us consider how the Hilbert action principle modifies in
empty spacetime.

Picture an empty region. Consider a functional variation of
the metric $\delta^{\rm f} g^{\mu \nu}$ in that region.
Observer stationed outside of the region who performs
coordinatization by using the same (say, radar) method as
before the variation, will find a slight change in the
coordinates of the marker events.

 Let us assume that such a change is proportional to the
functional variation in the metric via
\BEq
x^\alpha \rightarrow x^\alpha + \delta^{\rm f}
x^\alpha:= x^\alpha -
\frac{1}{2}\int_{O}^{P(x)} dx'^\lambda \;
\theta^\alpha_{\lambda\mu
\nu}(x')\,\delta^{\rm f} g^{\mu
\nu}(x'),
\EEq
where $\theta^{\alpha}_{\lambda\mu \nu}(x)$ is a
fundamental compensating tensor field, the {\it
coordinate compensator},
symmetric with respect to
$\mu\leftrightarrow\nu$ interchange. Integration is
performed along the unique geodesic that connects $P$ to some
{\it fixed} reference point $O$. The rare case of
several such geodesics -- the gravitational lensing --
usually corresponds to some sort of coordinate
singularity (for example, in the radar coordinatization
procedure less than four radar stations are needed to
pinpoint the focusing event).

When the Hilbert action
\BEq
S=\frac{1}{\kappa^2}\int d^4x \sqrt{-g}R
\EEq
is varied, there appears an extra contribution due to
the Jacobian of the coordinate transformation,
\BEq
\label{eq:ACTIONVARIATION}
\delta^{\rm f} S=\frac{1}{\kappa^2}\int \left((\delta^{\rm
f} d^4x)\;
\sqrt{-g}R +d^4x \; \delta^{\rm f} (\sqrt{-g}R) \right).
\EEq
We have:
\BEq
\frac{\partial (x^\alpha+\delta^{\rm f} x^\alpha)}{\partial
x^\nu}=\delta^\alpha_\nu+\partial_\nu (\delta^{\rm f}
x^\alpha),
\EEq
and 
\BEq
J\left(\frac{x+\delta^{\rm f} x}{x}\right)\equiv{\rm
det}\left(\frac{\partial (x^\alpha+\delta^{\rm f}
x^\alpha)}{\partial x^\nu}\right)
= 1 + \partial_\alpha \,(\delta^{\rm f} x^\alpha) = 
1-\frac{1}{2}\, \theta^\alpha_{\alpha\mu
\nu}(x)\,\delta^{\rm f} g^{\mu
\nu}(x),
\EEq
where we have used Mandelstam's definition for the
derivative of the line integral \cite{MAND62}.
Therefore the first term in
(\ref{eq:ACTIONVARIATION}) can be written as
\BEq
\label{eq:VAR1}
(\delta^{\rm f} d^4x)\;
\sqrt{-g}R=\frac{1}{2}\, d^4x\;\sqrt{-g}R\;
\theta_{\mu
\nu}\,\delta^{\rm f} g^{\mu
\nu}, \quad \theta_{\mu
\nu}\equiv \theta^\alpha_{\alpha\mu
\nu}.
\EEq
Varying the second term in (\ref{eq:ACTIONVARIATION}) gives
\BEq
\label{eq:VAR2}
d^4x \;\delta^{\rm f} (\sqrt{-g}R)=d^4x
\;\sqrt{-g}\left(-R^{\mu
\nu}+ \frac{1}{2}g^{\mu\nu}R\right)\delta^{\rm f} g_{\mu
\nu}.
\EEq
Combining (\ref{eq:VAR2}) with (\ref{eq:VAR1}) produces the
field equations
\BEq
R_{\mu\nu}-\frac{1}{2}\left( g_{\mu\nu}+
\theta_{\mu
\nu}\right) R
=0.
\EEq
Einstein's theory is trivially reproduced in the limit
$\theta^{\alpha}_{\lambda \mu \nu}=0$, but regardless of
compensator's value, the field equations have a flat
spacetime solution $R_{\mu \nu}=0$.

Finally, covariant continuity of the Einstein
tensor imposes four additional constraints
\BEq
\nabla^\mu\,
\left(\theta_{\mu\nu} R\right)=0
\EEq
on the ten components of $\theta_{\mu
\nu}$. 

\section*{Acknowledgments}

The author thanks David Finkelstein, Lewis Ryder, and Tony
Smith for encouragement and stimulating correspondence.

\end{document}